# Spreading of triboelectrically charged granular matter

**Deepak Kumar[1] , A. Sane[1], S. Gohil[1], P.R. Bandaru[1,2], S. Bhattacharya[1] and Shankar Ghosh*[1]**

[1] Department of Condensed Matter Physics and Materials Science, Tata Institute of Fundamental Research, Homi Bhabha Road, Mumbai 400-005, India
[2] Program in Materials Science, Department of Mechanical and Aerospace Engineering, University of California, San Diego, La Jolla, California 92093, USA

We report on the spreading of triboelectrically charged glass particles on an oppositely charged surface of a plastic cylindrical container in the presence of a constant mechanical agitation. The particles spread via sticking, as a monolayer on the cylinder's surface. Continued agitation initiates a sequence of instabilities of this monolayer, which first forms periodic wavy-stripe-shaped transverse density modulation in the monolayer and then ejects narrow and long particle-jets from the tips of these stripes. These jets finally coalesce laterally to form a homogeneous spreading front that is layered along the spreading direction. These remarkable growth patterns are related to a time evolving frictional drag between the moving charged glass particles and the countercharges on the plastic container. The results provide insight into the multiscale time-dependent tribolelectric processes and motivates further investigation into the microscopic causes of these macroscopic dynamical instabilities and spatial structures.

## Introduction

Granular materials often acquire charge due to triboelectric processes[1,2,3,4]. The resulting effect is seen in a wide variety of phenomena ranging from the very commonplace observation of food grains sticking on to their containers to problems of practical importance, e.g., sticking of pharmaceutical powders to tablet presses [5,6], particles sticking to the walls of pneumatic transport lines[7] and adhesion of toner particles in electrophotography[8]. The complexities associated are diverse. At a single particle scale, they are manifested through the microscopic charge transfer mechanism[1,2,3,4] and at larger scales, the collective response of the particles is related to the long range electrostatic interactions[5], that can lead to clumping[6] and pattern formation[7].

The motion of these particles on a substrate in response to external forcing is governed by the particle-substrate frictional interactions which relate to both the dissipative and the adhesive parts of the particle-substrate interaction[8]. In this paper we find that constant mechanical agitation modifies this frictional interaction over time. This is reflected in both the positional rearrangement and the motion of the particles on the substrate.

## Experimental setup

An orbital shaker from Heidolph Instruments produces triboelectric charging in a cylindrical container (vial) made of polypropylene (PP) or polystyrene (PS) (diameter $2R = 16mm$) containing $10^4$ glass particles. The shaker makes the vial move in a circle (radius $a = 2.5\ mm$ and at a frequency $f$) without spinning about its axis (**Fig. 1A** bottom panel). The spreading phenomena observed in both PP and PS vials are qualitatively similar (**Supplementary Movie SM1**). The specific vial for which the data are presented is mentioned in the text and figure captions. We use three kinds of imaging: (i) *continuous* (ii) *dynamic and (iii) static* for observing different aspects of the kinetics of the system. The continuous images are obtained using a fast camera (Phantom M310). The other images, i.e., dynamic and static are obtained using the camera EHD UK-1157. In the dynamic case, the system is imaged stroboscopically while in motion, at a specific location in its orbit. This is achieved by periodically triggering the camera with an electrical pulse generated by a position sensor (TCRT 5000). In contrast, the *static images* are obtained by temporarily pausing the drive and bringing the system to rest, following which the drive is resumed. The specific mode of imaging is explicitly mentioned in the text as well in the relevant figure captions. The orbital motion produces a potential well whose minima moves azimuthally along the surface of the cylinder in a periodic manner. This is reflected in the motion of the granular heap (see **Fig. 1 A** top panel), as observed by continuous imaging. The average charge acquired by the glass particles in the experiment is measured by pouring the particles into a Faraday cup connected to a Keithley 6514 system electrometer.

## Results

The constant rubbing of the heap against the vial surface leaves both the glass particles and the vial surface triboelectrically charged. For the present experiments, we find that rubbing leaves the glass particles with an overall positive charge (measured average charge per particle ~ $10^{-13}$ $C$) while the surface of the vial is oppositely (negatively) charged. The time evolution of this differential charging modifies the particle-substrate interaction. This causes a fraction of the particles to “stick to the wall” of the vial, albeit moving with a velocity less than $2\pi Rf$. The remaining particles continue to move as a heap with a linear velocity of $2\pi Rf$, i.e. the heap returns to its initial position after a complete period.

We present our findings in the following sequence. First, we discuss the structural arrangements of the stuck particles both with and without the drive. These arrangements change with time via the spreading of particles on the substrate, which, in turn, redistributes the counter charges (ions). Subsequently, we discuss the migration kinetics of the particle assembly, its relaxation processes and then analyze the motion of a single stuck particle. Finally, we present our results on the in-situ electrical measurements which provide additional information about the charges and their relative motion with respect to the vial.

**Overview of the spreading phenomena:** The sequence of dynamic images in **Fig. 1B** (**Supplementary Movie**: SM1 (Left panel)) captures the evolution of this heap with time ($t_w$) in a PP vial. As particles progressively stick to the surface, their number density spreads in the azimuthal direction and the shape of the heap (marked by lines in white) begins to change. From the dynamic images we compute the variation in the number density of particles with height ($h$): $\langle n(h)\rangle = \frac{1}{LI_{do}}\sum_{i=1}^{L} I_d(i,h)$. The angular brackets represent an average intensity of the dynamic image, $I_d(i,h)$, over the lateral dimension ($L$) of the image and $I_{d0}$ corresponds to the intensity of a single particle. The evolution of $< n(h) >$ with $t_w$ (**Fig. 1C)** shows two distinct regimes of spreading, viz., an initial phase (regime ***I***) marked by a slower increase of the spreading front, followed by a more rapidly changing front in the second phase (regime ***II***) that initiates after a characteristic time ($t_w^c$). The plateau height $h_{in}$ of the front in regime ***I*** scales linearly with the square of the drive frequency ($f^2$), i.e., the force applied, is independent of the particle diameter ($d$) and the material of the vial. Further, it is found to be approximately the same as that obtained with an equal volume of water replacing the glass particles, as shown in **Fig. 1D.**

As more particles stick onto the surface, a dense and amorphous monolayer coverage of the vial surface up to height $h_{in}$ builds up (see **Fig. 1B, Supplementary Movies SM1, SM3**) and remains nearly saturated for a long time interval. This saturated state in regime ***I*** can be thought of as a *composite system* consisting of stuck particles and their counter charges on the plastic surface. It is from this state in regime ***I*** that a new regime ***II*** evolves, marked by a further and rapid growth in the front height, a seemingly hithertofore unknown phenomenon which is, however, entirely absent in the same experiment performed on conventional liquids **(Supplementary Information: Fig. S1**). The crossover time ($t_w^c$) between regimes ***I*** and ***II*** depends inversely on the square of frequency, i.e., $t_w^c \sim 1/f^2$ as well as the material properties of the vial (**Fig. 1E**). For exampletriboelectric charging is proportional to the difference in the work functions ($w_F$) of the materials involved[9]. This is consistent with our observation that the crossover time for glass ($w_F$ =5 eV)[9] particles in polystyrene ($w_F$=4.2 eV)[10] vials is shorter than that in polypropylene vials ($w_F$=4.9 eV)[9]. The data presented in the paper are also sensitive to the humidity in the atmosphere[11] and the protocol used for cleaning the surfaces[12] implying the importance of the surface charges in the observed phenomenon.

***Transverse stripes - precursor to regime II:*** The onset of regime ***II*** is preceded by distinct changes in the structure of the amorphous monolayer of regime ***I***. The *dynamic* (top panel) and *static* (bottom panel) images in **Fig. 2A** (**Supplementary Movie SM2**) show the appearance of a periodic modulation in particle density in the azimuthal direction (transverse stripes) as observed in a PP vial. The angular distance between the stripes ($\Lambda$), *increases* with $f$ (**Fig. 2B**) and $d$ (**Fig. 2B** inset). The stripes are presumably accompanied by the formation of a similarly periodic spatial modulation of counter-charge density on the surface of the vial. This makes the modulation temporally robust and therefore observed in the static images too, as shown in **Fig. 2A** (bottom panel). The symmetry in the problem, a closed surface with periodic boundary conditions, preferentially supports an integral number of periods of this transverse density modulation. Indeed, for frequencies 14, 30 Hz (**Fig. 2B**) we observe more pronounced stripes which remain positionally locked in time while for intermediate values of $f$ (20, 25 and 40 Hz in **Fig. 2B**) the stripes are seen to slide and become substantially weaker in amplitude.

***Vertical streaks - onset of regime II:*** The formation of regime ***II*** is initiated by the emergence of a few-particle-wide vertical *streaks* that always originate from the tip of the stripes (see the image corresponding to $t_w = 7000s$ in the top panel of **Fig. 2A**). The large number density of particles in the stripe presumably localizes the charge and thus enhances the emanating electric fields leading to a breakdown-like flow of both particle- and charge-currents which initiates regime ***II***. At long times, however, these streaks gradually diffuse in the transverse direction and eventually merge to form a uniform growing front, as shown in the fourth panel ($t_w = 9000s$) in **Fig 2A**.

***Contrast in the structure of the stuck monolayer in the two regimes:*** In addition to the large change in the front velocity, the two regimes are very different in terms of their growth kinetics and structure. **Fig. 3A** and **B** show a sequence of time-lapsed images of the growth in regimes ***I*** and ***II,*** respectively) in a PP vial. In regime ***I***, the monolayer is formed by random adsorption of particles to the surface of the vial (**Supplementary Movie SM3**). The monolayer in regime ***II,*** in contrast, evolves by sequential adsorption of particles and a layer by layer growth of the spreading front (**Supplementary Movie SM4**). Furthermore, regimes ***I*** and ***II***, yield disordered and layered structures, respectively. Two-dimensional Fast Fourier Transform (2DFFT) images of the two cases are shown, respectively, in the left and right inset of **Fig. 3C**. The variation of $q_z$ and $q_\psi$, the wave vectors corresponding to the first peak of the 2DFFT along the axial and circumferential directions respectively, is shown in **Fig. 3C**. In regime ***I,*** (i) the system is isotropic, i.e., $q_z \simeq q_\psi$ and (ii) both $q_z$ and $q_\psi$ increase logarithmically in time. But in regime ***II***, (i) the structure near the spreading front is anisotropic, $q_z < q_\psi$, and exhibits periodicity in the axial direction (regions of localized 'spots' in the Fourier images) and (ii) $q_z$ and $q_\psi$ remain constant in time. We note that such structural anisotropy has been considered previously in driven disordered media[13].

**Migration kinetics of the particle in the two regimes:** The particle motion in the two regimes depends on the distribution of charge on the substrate which provides the frictional interactions for the particles. To explore their spatial nature, we analyse images from a region closely below the spreading front and examine the following:

(1) *The summed image*: $[\sum_{t_w} I_s(t_w)]$ – a superposed sum of the static images, $I_s(t_w)$**.** This image captures the different realizations of the configurations of the stuck particles and therefore closely represents the energy landscape associated with the charge distribution on the surface of the vial. The summed image has no distinct structure barring for a hint of granularity in regime ***I*** (**Fig. 3D** left panel), i.e., the stuck particles are sliding in the presence of the drive. In regime ***II***, (**Fig. 3E** left panel) it has a layered structure with a well-defined periodicity along the vertical axis with no modulation in the azimuthal direction. Since the mobility of the particle in the vertical direction is hindered, we infer the frictional interaction in regime ***II*** to be anisotropic (small along the azimuthal direction and large in the axial direction). This is further substantiated from the difference images described below.

(2) *The difference image*: $[|I_s(t_w) - I_s(t_w + \Delta t_w)|]$, is taken between two static images separated by a time difference $\Delta t_w = 100s$. The bright and dark areas of the difference image below the spreading line correspond to the spatial regions associated with mobile and stagnant particles respectively. For these images, the bright spots are homogenously distributed in regime ***I*** (**Fig. 3D** middle and right panels), while they appear localized in regime ***II*** (**Fig. 3E** middle and right panels). Thus, in regime ***I***, the particles can move isotropically while in regime ***II*** particles whose axial motion is restricted coexist with those which can hop between layers. This is also further evidenced by a rougher spreading front in regime ***I*** compared to regime ***II***.

(3) *The superposed difference image*: $[\sum_{t_w} |I_s(t_w) - I_s(t_w + \Delta t_w)|]$. The superposed difference image (right panel of **Fig. 3E**) identifies the hopping paths across the layers in regime ***II*** through

which the particles advect upwards and then sequentially deposit at the top. The averaging over different realizations of configurations ensures that the bright regions in these images correspond to the localised spatial tracks of particles that hop. This form of particle diffusion is characteristic of smectic ordering[14]. Such localized transport of particles are, however, absent in regime ***I*** (right panel of **Fig. 3D).**

**Relaxation dynamics:** We now discuss the temporal evolution of the structure of the monolayer of the stuck particles after the drive is switched off. At short times (small values of $t_w$) the particles spread azimuthally and at long times they detach from the surface. These results highlight the role of thermal processes , e.g., diffusion of charge carriers, in the otherwise athermal granular system.

The short time azimuthal spreading happens via a number of transient structures. *e.g.*, chain-like formations (**Fig. 4A**) in the PS vials and are characterized by the temporal evolution of their average chain length, $l_{chain}$ (**Fig. 4C, Supplementary Movie SM5**) in regime ***I***. These chains represent the local patchiness of the charge distribution[15] on the plastic surface. One observes multiple time scales in this spreading, e.g., the system initially spreads azimuthally within a few seconds. It then starts to arrange itself in the form of vertical chains, whose average length reaches the maximum value in about 10 $s$. For the next 100 $s$ the chains spontaneously break and $l_{chain} \propto \frac{1}{\sqrt{t}}$. If one assumes that the length of the chain is proportional to the depth of the effective potential that stabilizes it, then its inverse square root dependence on time would result from the reduction in this depth due to diffusive dynamics of the countercharges on the substrate[16].

We now discuss the stability of the monolayer of particles stuck on the PP vial upon the cessation of driving. While the particles close to the top are well-adhered, the stability of those stuck at lower heights is found to depend on $t_w$. **Fig. 4B** (**Supplementary Movie: SM6**) shows a typical time-lapsed sequence of images showing particles falling off the vial's surface after the orbital motion is turned off. The number density of the stuck particles decays in a manner shown in **Fig 4D**. Here $N_s(t_w)$ is the number of stuck particles at the instant when the drive is switched off and $t$ is the time measured since then. At long times, the decay can be approximated to $N_s(t_w + t) = N_s(t_w)(t/\tau_g)^{-\alpha}$ (**Fig. 4D**). A power-law decay typically implies the existence of multiple relaxation processes and a hierarchical structure of the potential energy landscape associated with the particle substrate interaction. Here, $\tau_g$ is the mean of the life time distribution of the stuck particles and the exponent $\alpha$ is a measure of its width. The variation of $\log(N_s(t_w)/N_s(t_w+t))$ (which is a measure of $\alpha$) with $t_w$, is shown in **Fig. 4E** for $t = 10s$. Within this scheme of analysis, $\alpha$ is nearly constant in regime ***I***, while it varies substantially in regime ***II***. The mean life time ( $\tau_g$) is very large in regime ***I*** and decreases with $t_w$ (**Inset of Fig. 4D**) in regime ***II***. A visual inspection of the system, six months after the completion of the experiment, shows that the monolayer close to the top remains intact (**Supplementary Information: Figure S4** ).

**Single particle kinetics:** For small values of $t_w$ the particles form a heap which moves with a velocity $2\pi Rf$ as shown in **Fig.1A**. A fraction of the particles get stuck to the surface of the vial with increasing time. In this section we analyse the motion of stuck particles under the influence of the periodic forcing due to the orbital motion. Typical time traces of the angular position ($\Delta\theta$) of stuck particles at three different values of $f$ $(10, 12, 14\ Hz)$ are shown in **Fig. 5A**. It exhibits an oscillatory behaviour with an asymmetric forward and reverse motion. The trajectory of the particle on the surface of the vial for one period of the orbital motion is shown in **Fig. 5B (red line)**. The blue and green circles mark the position of the particle at the beginning and the end of the cycle, respectively. The initial configuration of the rest is marked by black dots. The angular shift between the blue and the green circles reflects its drift along the drive. The inset of **Fig5A** shows the probability distribution of the amplitude of the forward motion ($\theta_0$, marked in **Fig. 5A**) and the drift velocity ($v$), calculated from the time traces of hundreds of stuck particles driven at $f = 12Hz$. The velocity distribution is particularly broad and highly asymmetric.

The motion of a single particle under the periodic forcing is captured in a simple overdamped equation of motion: $\mathrm{R}\Gamma \frac{\mathrm{d}\theta}{dt} = m(2\pi f)^2 a \sin(\theta - 2\pi f t)$ where Γ is the measure of the frictional coupling. The term on the right hand side is the tangential component of the centrifugal force at the angular position ($\theta$) of the particle on the vial. In general, this coupling depends on the stress history of the system[17,18] and is typically viscoelastic[19], hence Γ is a complex quantity ($=\Gamma' + i\Gamma''$)[8] . Microscopically the real part of Γ is related to the dissipative mechanisms in the system.

This can arise due to inelastic inter-particle and substrate particle interactions. In the present scenario, the dissipation in the substrate-particle interactions is mainly due to the resistive loses associated with the ionic current in the vial. The momentum transferred from the sliding particles to the counter-charges produces this current. The spatial variation in the electrostatic potential between the particle and the substrate provides the elastic coupling of the particles to the local energy minima (pinning sites)[8] and is related to the imaginary part of Γ. The variation of $\theta$ as a function of $t$ is obtained by numerically solving the above equation. Since we observe that the particle moves over distances much larger than the typical interparticle separation (see **Fig. 5B**), we neglect the imaginary part of Γ in the numerical integration. In the limit $\frac{2\pi f m a}{\mathrm{R}\,\Gamma} > 1$ , at long time, $\theta \sim 2\pi f t$, which is representative of the motion of the heap (**Fig. 1A**). For $\frac{2\pi f m a}{\mathrm{R}\,\Gamma} < 1$ , $\theta$ oscillates between two extremes that drift linearly with increasing time. **Fig. 5C** shows $\Delta\theta(t)$ for $\frac{ma}{\mathrm{R}\,\Gamma} = \frac{1}{500} s/rad$ and $f = 10\ Hz, 14Hz.$ These time traces are similar to the ones obtained experimentally. Note, however, that it does not capture the inherent randomeness observed in the jagged experimental time traces. A simple way to incorporate the randomness in an effective single particle equation of motion is through a spatially varrying friction coefficient[20,21], which is amply justified by the large distribution of single particle velocities measured in the experiment (see inset of **Fig. 5 A**). The frequency dependence of the mean amplitude of the forward motion ($\theta_0$) and the drift velocity of the extremas ($\nu$) for the experimentally obtained traces is shown in **Fig. 5D** and **E**, respectively. The distribution in the drift velocity, $\nu$, is particularly broad and its variance is of the order of the mean. A threshold behaviour, typical of friction dominated dynamics[22], is particularly visible in the variation of $\nu$ with $f$. Frictional ageing in the system is reflected in the evolution of $\nu$ and $\theta_0$ with $t_w$ (for $t_w < t_w^c$) as shown in **Fig. 5 F** and **G,** respectively.

The local variation in the time evolution of the particle substrate frictional interaction aids the formation of the various growth patterns observed in the experiment. With time, the system spatially segregates into regions of high and low densities where particles have small and large mobilities, respectively. Futher, the frictional interactions evolve from being locally isotropic in **regime I** to being anisotropic in **regime II**. The former can be observed from the time averaged image shown in **Fig. 3D** first panel, while the latter is evident from the layered structure observed in the time averaged image in **Fig. 3E** first panel.

The real part of the frictional coupling (Γ) is related to the ionic diffusivity $\alpha$ in the system. This can be estimated from the growth of the particle front in the direction perpendicular to the mechanical drive. During the onset of ***regime II*** (**Fig. 1**) the front moves by a distance, $\ell \sim 10\ mm$, in a time interval, $\delta t_w \sim 10$ hours. From this we estimate a diffusion coefficient $\alpha \left(= \frac{\ell^2}{\delta t_w}\right)$ = 3× $10^{-3}$ $mm^2/s$. The values of $\alpha$ is found to increase with $f$.

**Electrical Measurements:** Along with imaging as described above, the charge developed in the system is simultaneously monitored by placing an electrode near the vial at a fixed location in the laboratory frame shown schematically in the **insets of Fig. 6A**. As discussed in **Fig.1,** the centre of the vial moves in a circular orbit. At the closest point of approach, the electrode is $s = 500\mu m$ away from the vial's surface. The current induced in the electrode depends on the local (near the electrode) uncompensated charge on the vial, $\delta Q$, and its velocity ($u$) with respect to the electrode, i.e., $i = \delta Q.u.\phi(r)$, where $\phi(r)$ is a proportionality constant whose value depends on the distance ($r$) between the electrode and the vial. In the present experimental geometry the distance between the centre of the vial and the electrode ($r(t) = \sqrt{A^2 + a^2 - 2\ a\ A\cos(2\pi f\ t)}$) and therefore the measured induced current ($i$) varies in an oscillatory manner; here $A(= a + R + s)$ is the distance between the electrode and the centre of the orbit. A *lock-in based phase sensitive technique* has been used to measure the amplitude $i_0$ and the phase ($\delta$) of the current $i$. The phase is measured with respect to the vial's motion in the orbit. We use the trigger pulses generated by the position sensor (TCRT 5000) as the reference signal for the lock-in measurements.

***Control experiments:*** To determine the dependence of $i_0$ and $\delta$ on the charge in the system, the frequency of the orbital motion ($f$) and the distance of closest approach ($s$) of the vial from the electrode, we perform the following control experiment. A conducting tube is inserted inside the vial so that it is in contact with it and a dc voltage ($V$) applied on the conducting tube. The charge on the conducting tube is proportional to the applied voltage. We find that the amplitude, $i_0$, varies linearly with $V$ **(Fig. 6A**) and frequency $f$ (**Fig. 6C**) and it decreases with $s$ (**inset of Fig. 6B** ). The conducting tube has a self-capacitance of $C = 20\ pF$, and the slope of the linear fit between $i_0$ and $V$ $(= charge/C)$ provides a proportionality constant $\sim (10^7 charges/cm^2 Hz)/nA)$ . The phase ($\delta$) is sensitive only to the sign of $V$ (**Fig. 6 B** and **D),** is independent of the magnitude of $f$ **(Fig. 6D**) and depends weakly on $s$ (**Fig. 6B inset**).

***In situ experimental results:*** The time evolution of $i_0$ and $\tan(\delta)$ for the experiment described in **Fig. 1** is shown in **Fig. 6E and F**, respectively. This method does not measure the net charge on the particle or the substrate. Instead, it is sensitive to the charge imbalance ($\delta Q$) at a local scale, i.e. near the electrode. The phase is sensitive to the motion of the charges with respect to the surface of the vial. A typical measured $i_0 {\sim} 1\ nA$ (**Fig. 6E**) corresponds to a net unbalanced charge of $10^7\ charges/cm^2 Hz$. This provides a lower bound of $10^{-14}\ C$ of charge per particle ($d = 300 \mu m$). Alternatively, the average charge acquired by the glass particles in the experiment is measured by pouring the particles into a Faraday cup connected to a Keithley 6514 system electrometer. This method provides a value of charge per particle ($d = 300 \mu m$) to be equal to $10^{-13}\ C$. Near the end of regime ***I***, one observes a rise in $i_0$ and a distinct cusp in $\tan(\delta)$ at a time which precedes $t_w^c$. These precursor events lead to the onset of regime $\boldsymbol{II}$**.**

The variation of $\frac{i_0}{f}$ and $\tan(\delta)$ with frequency for the experiment described in **Fig. 5D** and **E** is shown in **Fig. 5 H** and **I.** From the control experiments we expect $\frac{i_0}{f}$ to have a constant value. The change in the slope of these curves coincides with the onset of drift of the particles (**Fig. 5E**). Similar variation in $i_0$ and $\tan(\delta)$ is observed at a given frequency as a function of time as the system crosses over from **regime I** to **II (Fig. 6E and F),** presumably related to the observed collective migration of particles (**Fig. 1 C**).

## Conclusions

The present problem of an assembly of interacting particles being driven over a substrate is a particular realization of the more general problem of driven many body systems in a random potential which is commonly assumed to be time-invariant, i.e., quenched. In the present system the motion of the particles produces the triboelectric charging, thereby modifying the potential energy landscape over which the particles move. This, in turn, modifies the motion of the particles. This coupling between the spatial structure of the potential and the collective dynamics of the particles ensures that both evolve in a self-consistent manner, thereby producing novel self-assemblies that evolve in time. Time evolution of frictional coupling has been previously observed in a variety of systems, e.g., friction induced wear[22], frictional ageing in rocks[23], sticking of particles to substrates[8]. However, to our knowledge, the long-scale structural reorganization of the resulting potential energy landscape and their effects on the particle-assembly have been seldom investigated. We expect that the present set of observations will motivate further work on triboelectric phenomena as a rich source of pattern formation of particle assemblies moving over a time-evolving potential landscape.

**Contributions**

D.K and S.Ghosh conceived the experiments. D.K and S.Ghosh carried out the experiments and data analysis with the help of A.S and S.Gohil. D.K , S.Ghosh,P.B and SB contributed to the discussion of the results and writing of the manuscript. All authors reviewed the manuscript.

**Competing financial interests**

The authors declare no competing financial interests.

## Methods

### Protocol for cleaning:

To clean the soda-lime glass particles obtained from a local vendor, we heated them at 200°C for about an hour and then let them cool down to room temperature before starting the experiment. A soda-lime glass slide when treated in the same way shows that the contact angle made by a drop of water on it decreases significantly from 65° (**Supplementary Information: Fig. S2 A** left panel) to 14° (**Supplementary Information: Fig. S2 A** middle panel). When the contact angle measurement is repeated on the same slide after two days no significant deviation from 14° is observed (**Supplementary Information: Fig.S2 A** right panel). Thermogravimetric analysis (TGA) on the uncleaned glass particles shows a sharp dip in mass by 0.01% at $T = 169°C$ (**Supplementary Information Fig. S2 B** (gray curve)). On repeating the TGA measurement on the same particles, the feature is not observed (**Supplementary Information: Fig. S2 B** blue curve). This implies that the volatile impurities have been removed from the surface by heating.

The polystyrene vials used in the experiment were obtained from Griener Bio One (Cat.-No.: 188171) and the polypropylene vials from Tarsons (Cat No- 523070, Lot No – B280312). The vials have been cleaned using dilute HCl solution. The vials are kept dipped in an aqueous solution of HCl (100 mM) for 10 hours, subsequent to which they are extensively rinsed using deionized water. The vials are then left in the oven at 60°C for about 12 hours to dry. We observe that the given protocol of cleaning makes the surface of the vial smooth (**Supplementary Information: Fig S2 C**). Experiments have also been performed by increasing the time for which the vials are exposed to the acid by about 5 times (**Supplementary Information: Table T1, Fig. S3**).

**Relative humidity:** The humidity of the room where the experiments were carried out was maintained at ∼55% using two dehumidifiers (White Westinghouse Dehumidifier – WDE201).

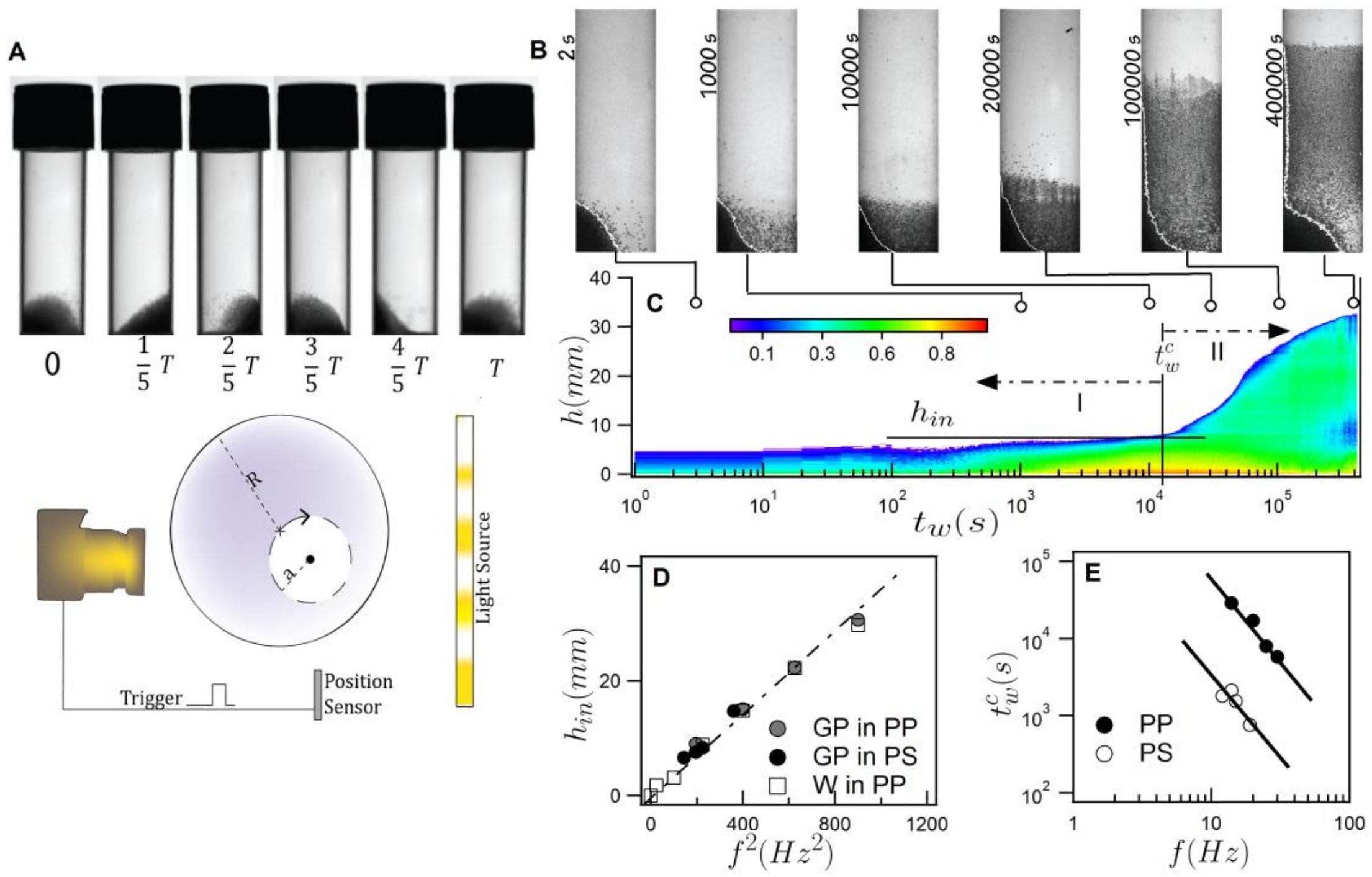

**Figure 1 Spreading dynamics.** (A) Bottom panel: The experimental setup consists of a cylindrical polypropylene vial of diameter ($2R = 16mm$) containing $10^4$ glass particles (average diameter $d = 300\ \mu m$, polydispersity = 20%, and average individual mass $m = 40\mu g$)) whose center (marked by a cross) moves on a circular orbit (dashed line) of diameter ($2a = 5mm$) at a constant frequency $f$. The orbital motion is achieved using an orbital shaker. (A) Top Panel: The granular heap as imaged in the continuous mode at six equally spaced time intervals in one period ($T = 1/f$). (B) Typical images (dynamic mode) of the vial (polypropylene) taken at certain instances of times (marked in the sequence), while it is in orbital motion at $f = 14Hz$. The lines in white outlines the heap of particles formed due to inertial force arising out of the orbital motion. (C) The evolution in time of the number density of particles $< n(h) >$ as a function of height. The numerical value of $< n(h) >$ is color-coded. The growth regimes ***I*** and ***II*** are marked. (D) The variation of $h_{in}$ as a function of the square of frequency of orbital motion for glass particles (GP) in polypropylene (PP) vial (gray circles), GP in polystyrene (PS) vial (black circles) and water (W) in PP (open squares). (E) The variation of $t_w^c$ as a function of frequency for glass particles in polypropylene (PP, black circles) and polystyrene (PS, open circles) vials. The solid line has a slope of -2 in the log-log axis, i.e., $t_w^c \sim f^{-2}$.

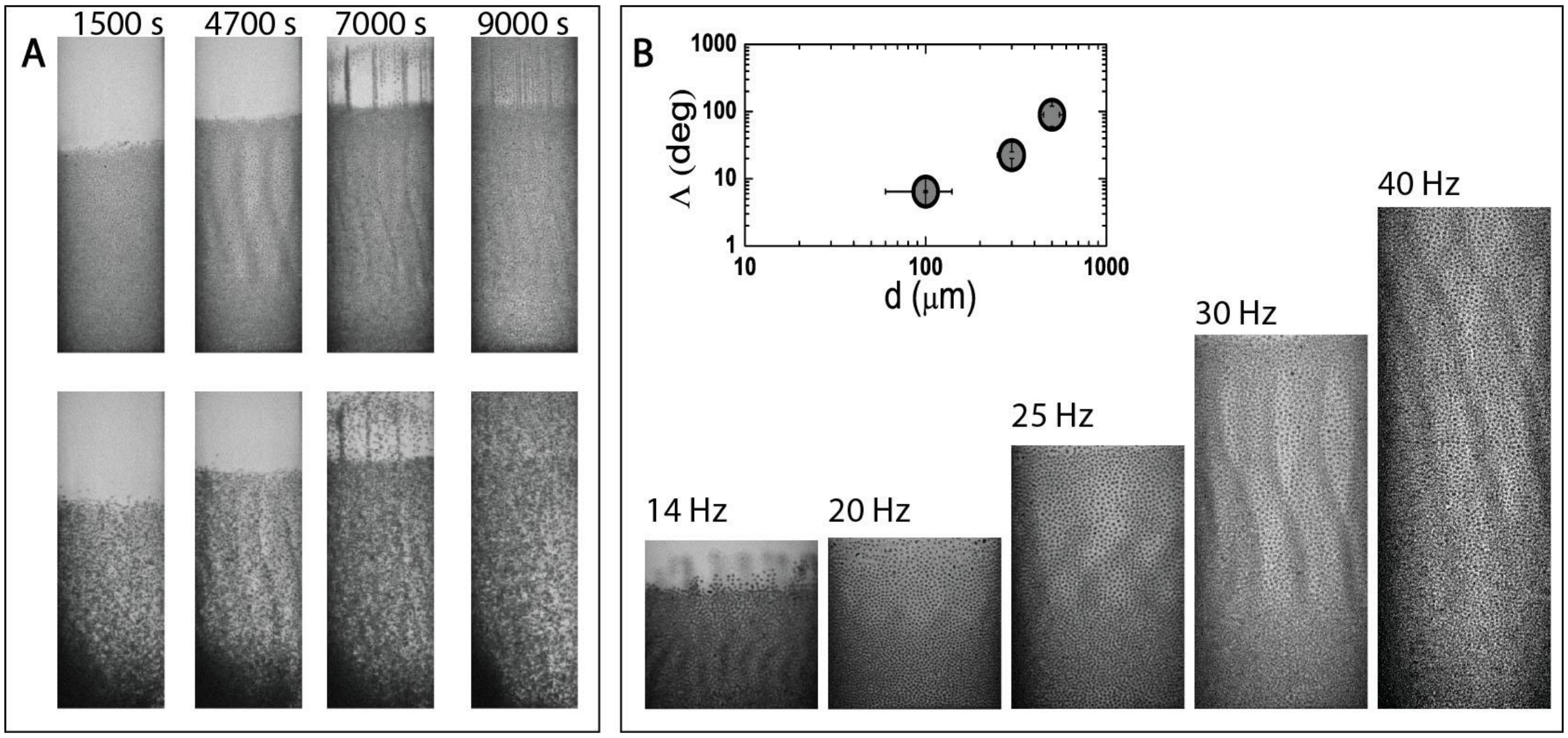

Figure 2 **Transverse density modulation.** (A) The appearance of a modulation in the density of stuck particles ($d$=$300\mu m$) with a periodicity in the azimuthal angle in polypropylene vial. Top and bottom panel shows static and dynamic images at a sequence of times marked above the images. The periodicity of the modulation depends on the frequency of the orbital motion and the size of the particles. (B) shows this periodic modulation at frequencies 14Hz, 20Hz, 25Hz, 30Hz and 40 Hz. For certain frequencies (14 Hz, 30 Hz) we observe more pronounced stripes which weaken substantially for intermediate values of $f$ (20 Hz, 25 Hz and 40 Hz). Inset shows the dependence of the wavelength of the periodic modulation, Λ ,on the particle diameter, d, at $f = 14Hz$.

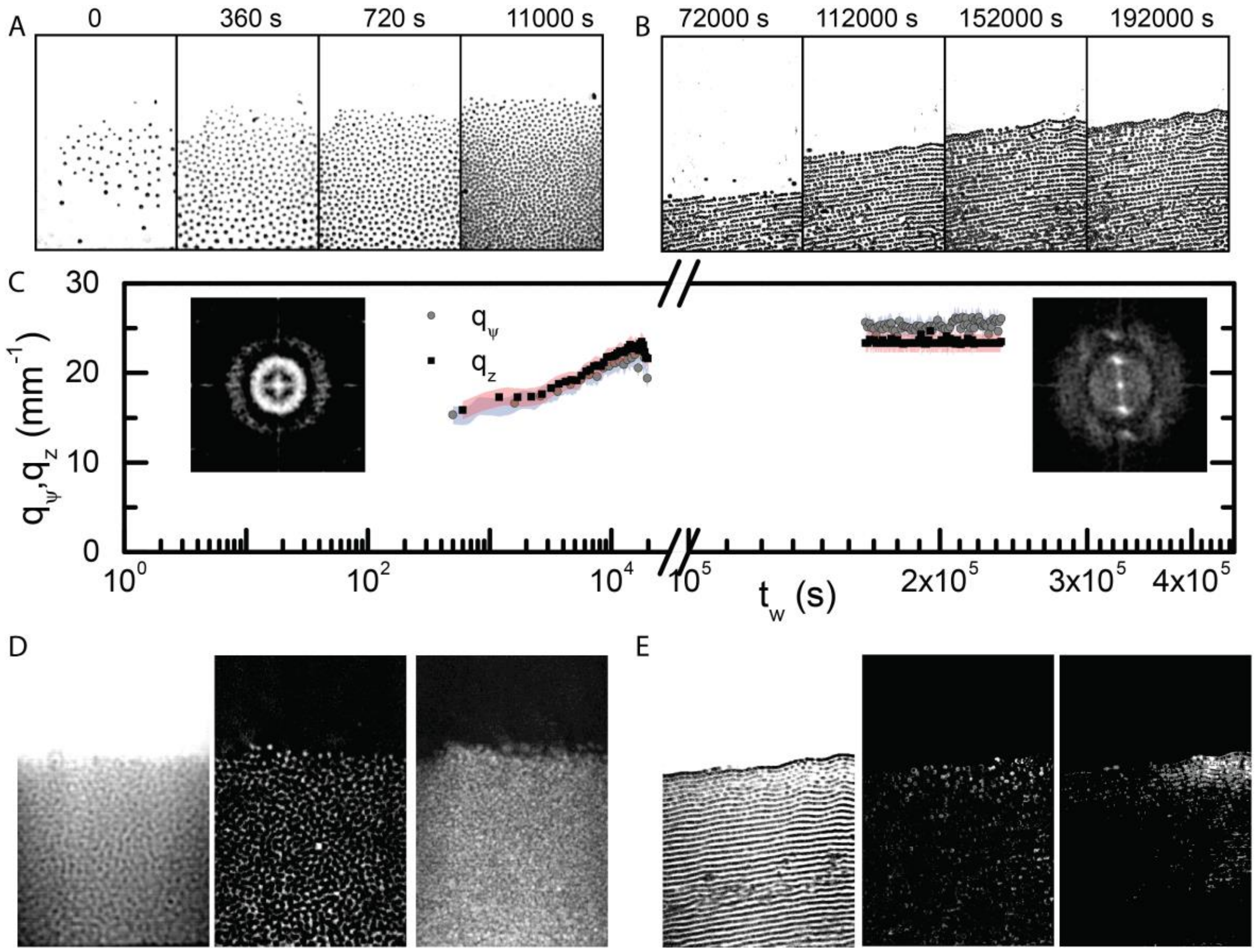

Figure 3 **Comparison of the growth at small and large times.** In regime ***I*** the system grows through random adsorption of particles on the surface of a polypropylene vial. This is shown in the images (static mode) in A. In the later part of regime ***II***, layer by layer growth takes over as is reflected in the sequence in B. (C) The $q_z$ (square) and $q_\psi$ (circle) corresponding to the first peak in the 2DFFT of the structure in regime ***I*** and regime ***II*** as a function of time. The confidence interval of the data is marked by translucent bands. Left and right insets are examples of 2DFFT obtained in regime ***I*** and ***II*** respectively. The average of 50 static images ranging over a time $\sim 5000\ s$, difference between two consecutive static images separated by 100s and the average of such subtracted images over a time period of $\sim 5000\ s$ are shown in the left, middle and right panels for regime ***I*** in D and regime ***II*** in E, respectively.

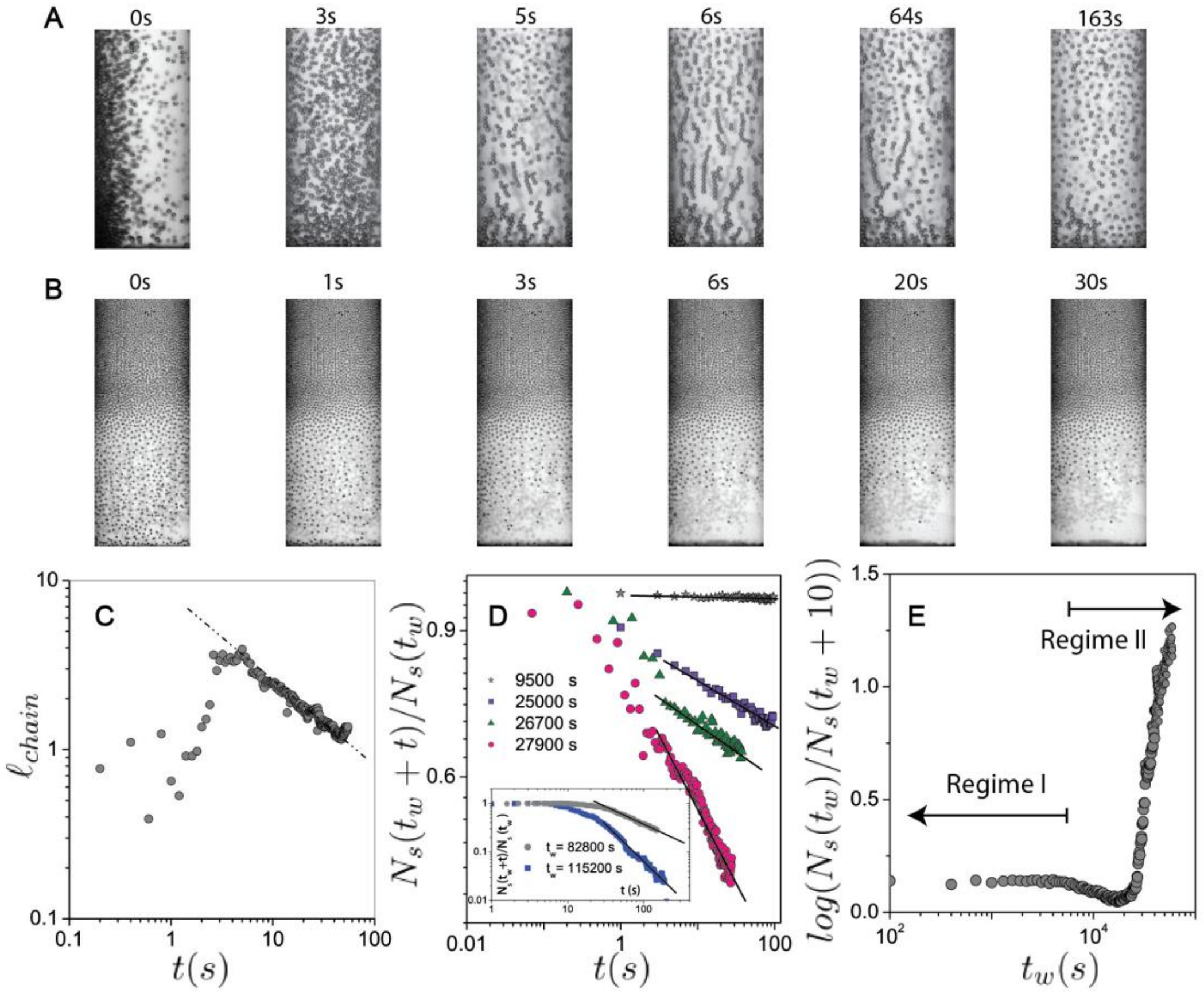

Figure 4 **Time evolution on stopping the drive.** (A) For $t_w < t_w^c$ (Regime ***I***), the initial state consists of particles concentrated in a high density region at one azimuthal position as in the first panel of A. In the final state long after the motion has stopped, the particles are uniformly spread over the entire surface with roughly the same inter-particle separations as in A last panel. The transition from the initial to the final configuration goes through an intermediate state where long chain like structures are favored. (C) shows the evolution of the average chain length with time $t$. For large values of $t$, $l_{chain} \propto \frac{1}{\sqrt{t}}$ (shown by the dotted line). The data shown in A and C is for $500\ \mu m$ glass particles in polystyrene vial. (B) For $t_w > t_w^c$ (Regime ***II***), the particles stuck at lower height fall down and the density of stuck particles decreases and eventually becomes zero. This data is for $d = 300\mu m$, $f = 25\ Hz$ and $t_w = 2.7 \times 10^4\ s$ in a PP vial. The change in the density of stuck particles as a function of time is shown in D for different values of $t_w$ (marked in the figure). In the inset of (D) we show the same variation for another system consisting of $500\mu m$ particles. At large times we observe a power-law decay (shown by the solid lines), whose exponent, $\alpha$, varies as a function of $t_w$. (E) shows the variation of $\log(N_s(t_w)/N_s(t_w + t))$, which is a measure of $\alpha$, with $t_w$ for $t = 10s$. The data shown in B, D and E is for $300\ \mu m$ glass particles in polypropylene vial.

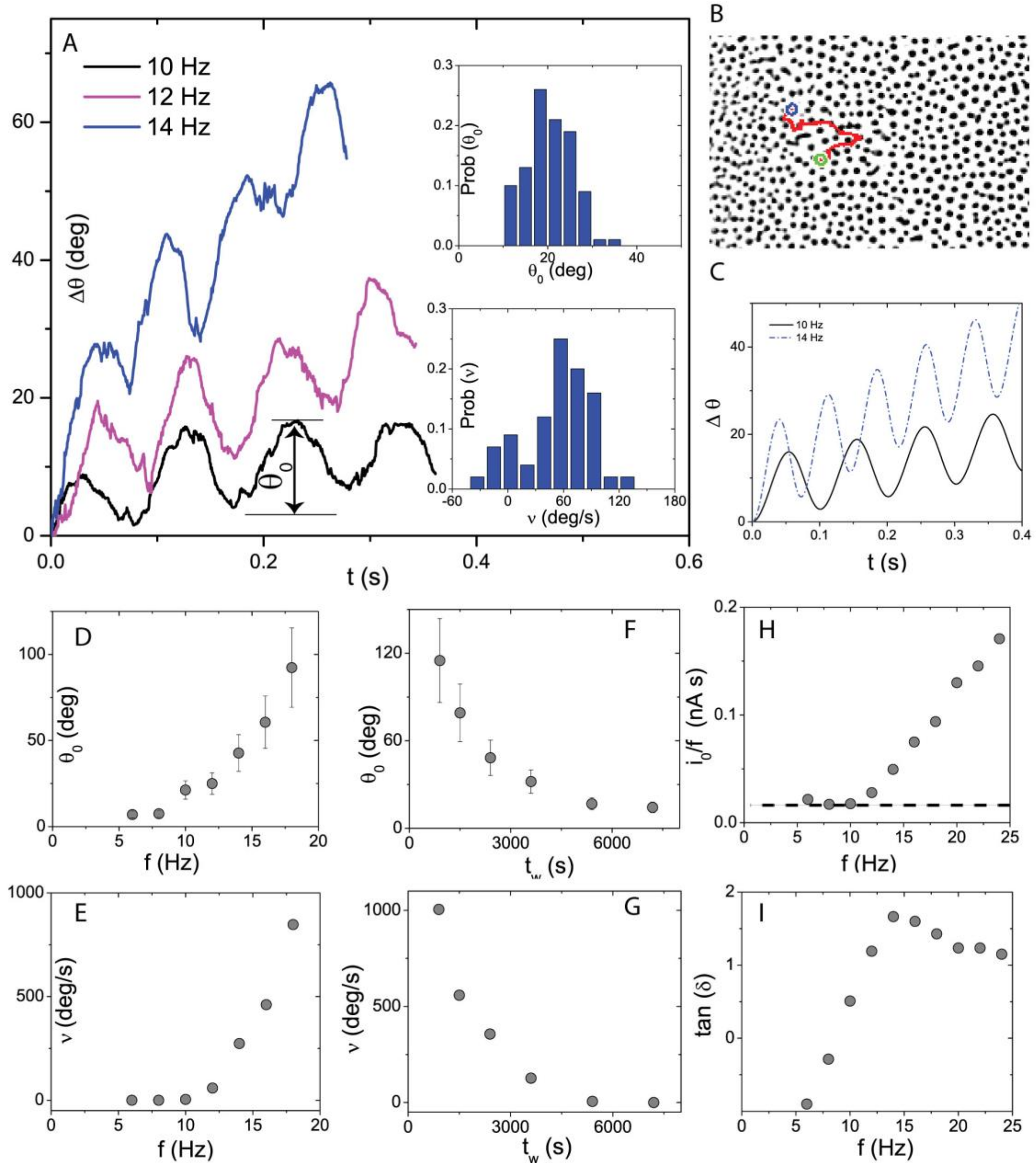

Figure 5 **Single particle kinetics.** (A**)** Typical time traces of the angular position of stuck particles ($d = 300\mu m$) at three different values of $f$ measured at a given value of $t_w \ll t_w^c$. The two insets show the distribution in the amplitude of the forward motion, ($\theta_0$) and the drift velocity ($\nu$) obtained over hundreds of stuck particles at $f = 12\ Hz$. (B) A typical trajectory of stuck particle (red line) during one complete cycle of orbital motion. The position of the particle at the beginning and the end of the cycle is marked by blue and green circles respectively. The black dots in the background show the positions of the rest of the particles at the beginning of the cycle. (C) Time trace of the angular position obtained using the model described in the text for $f = 10\ and\ 14\ Hz$, and $\frac{ma}{R\Gamma} = \frac{1}{500} s/rad$. Variation of $\theta_0$ and $\nu$ with $f$ is shown in (D) and (E) respectively. (F) and (G) show the evolution of $\theta_0$ and $\nu$ with $t_w$, respectively, for $t_w < t_w^c$. Typically $\nu$ is observed to have a variance of the order of the mean values plotted and is not marked in the plots for clarity. (H) $\frac{i_0}{f}$ and (I) $\tan(\delta)$ as a function of $f$. In (H) the dash-dot line marks an $i_0$ varrying linearly with $f$, the behaviour seen in the control experiment described in **Fig.6C**.

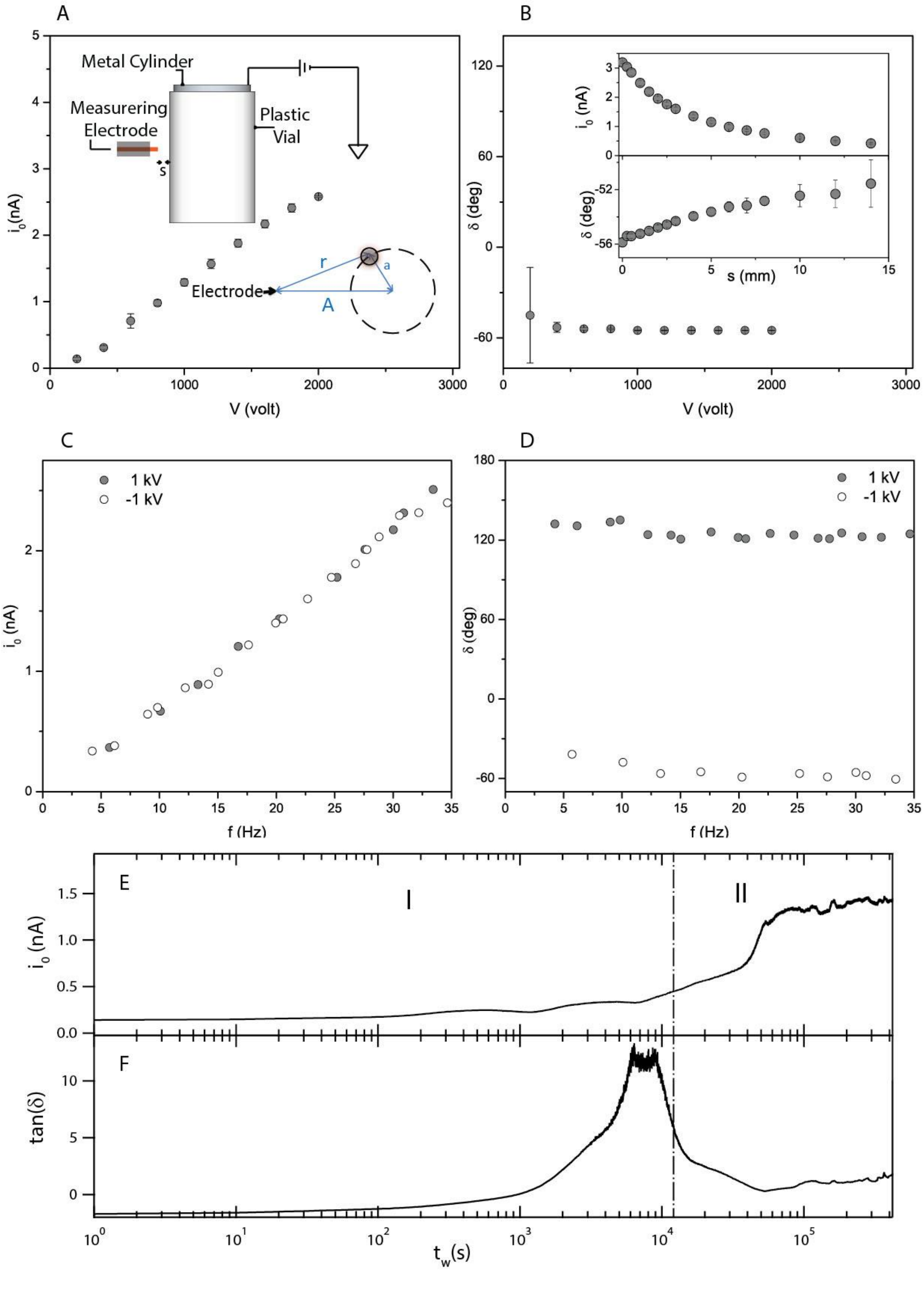

Figure 6 **In-situ electrical measurements.** The insets show the schematic side and plan view of the electrical measurement setup. In the plan view, the vial (marked by solid circle) moves on a circular orbit (dashed circle). The distances $A$ and $r$ mark the distance of the electrode from the centre of the orbit and the instantaneous position of the centre of the vial respectively. The distance of closest approach from the surface of the vial to the electrode is shown by $s$. Control experiment on the electrical measurement setup is performed by inserting a conducting tube inside the PP vial, which is connected to high voltage source. (A) and (B) show the measured amplitude and phase as a function of the voltage applied on the conducting tube. In this experiment, $f = 20\ Hz$. The inset of (B) shows the variation of $i_0$ and $\delta$ with the closest distance between the measuring electrode and the vial. (C) and (D) show $i_0$ and $\delta$ as a function of $f$ for two values of applied voltage $V = 1kV\ and - 1kV$. The presence of a conducting tube to which a voltage is applied is specific to the control experiment. This is absent in the main experiments (described in Fig.1). (E) and (F) show the simultaneously measured $i_0$ and $\delta$ during the experiment described in Fig. 1. The dash dot line marks the position of $t_w^c$.

# Supplementary Information

## Movie Captions

**Movie SM1:** (Left) A typical experiment with glass particles in a Polypropylene vial. Each individual time stamped frame is a dynamic image taken while the system is under orbital motion at f=14 Hz. (Right) A typical experiment with glass particles in Polystyrene vial at f=18 Hz, consisting of dynamic images. The scale bar is 2mm.

**Movie SM2:** Appearance of the transverse modulation (stripes) in particle density. This is a sequence of static images taken while the system is brought temporarily to rest by suspending the drive during the experiment. For this experiment, f=30 Hz. The scale bar is 2mm.

**Movie SM3:** Formation of regime I as seen from the static images. The number density of particles increases with time as more particles get adsorbed on the surface randomly. The scale bar is 2mm.

**Movie SM4:** Formation of regime II at large times. The scale bar is 2mm.

**Movie SM5:** Time evolution of the structure as the orbital motion is switched off and the vial becomes stationary in the laboratory frame. The movie corresponds to regime I, for 500 μm glass particles in a Polystyrene vial. The particles which are concentrated towards one end due to the orbital motion of the vial, spread out to form uniform coverage. However, it goes through an intermediate state where chain like structures oriented in the vertical direction form and then spontaneously break-up. The scale bar is 2mm.

**Movie SM6:** Time evolution on switching off the orbital motion for $t_w \gg t_w^c$. The particles which are stuck on the lower part of the wall during the orbital motion, gradually fall down under the influence of gravity. The experiment shown corresponds to 500μm glass particles in a Polypropylene vial driven at 14 Hz. The scale bar is 2mm.

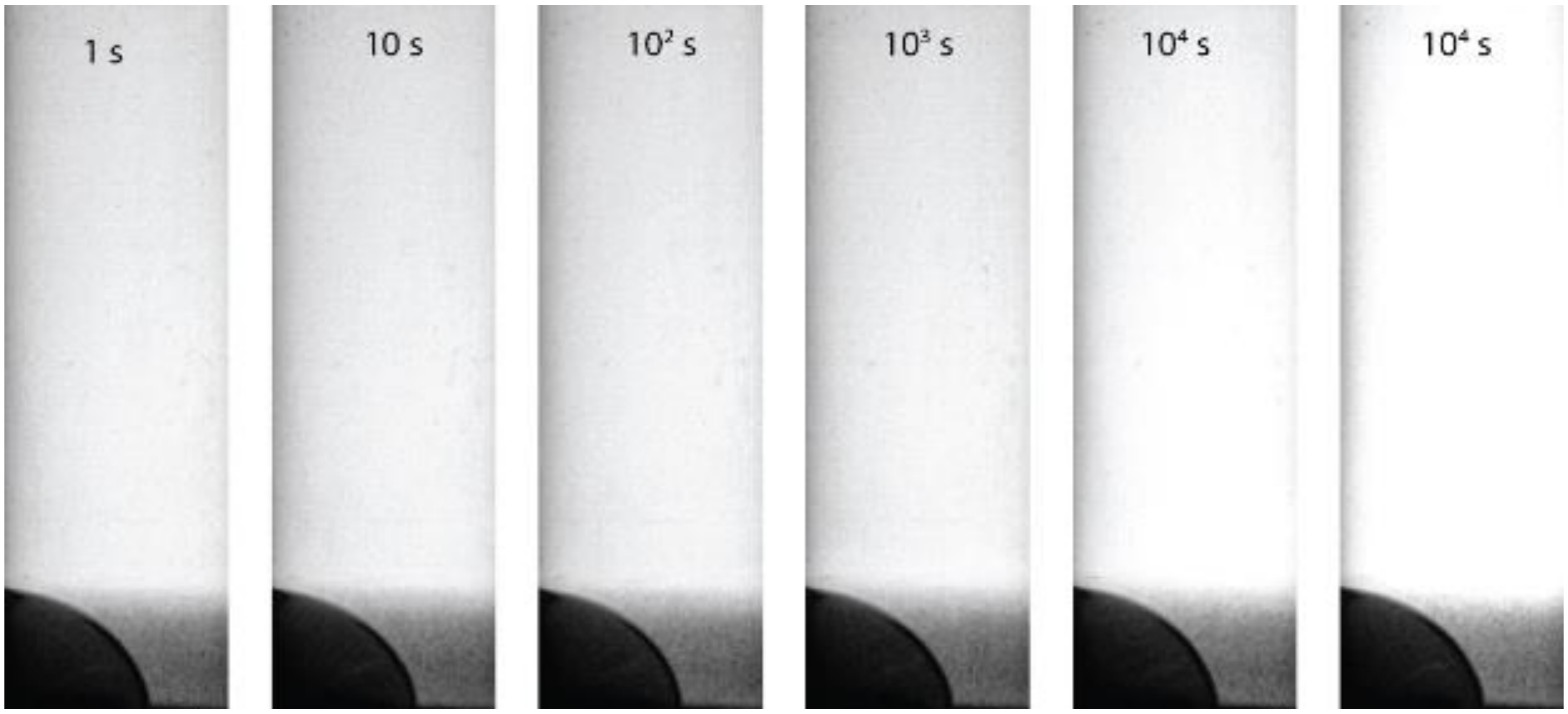

**Figure S1** The result of a “control” experiment performed on Silicone oil (viscosity 10 cSt) in a Polypropylene vial at f=14 Hz. The sequence of dynamic images taken at times ($t_w$) are marked. Note that all the images appear identical, i.e., the system shows no signs of evolving with time.

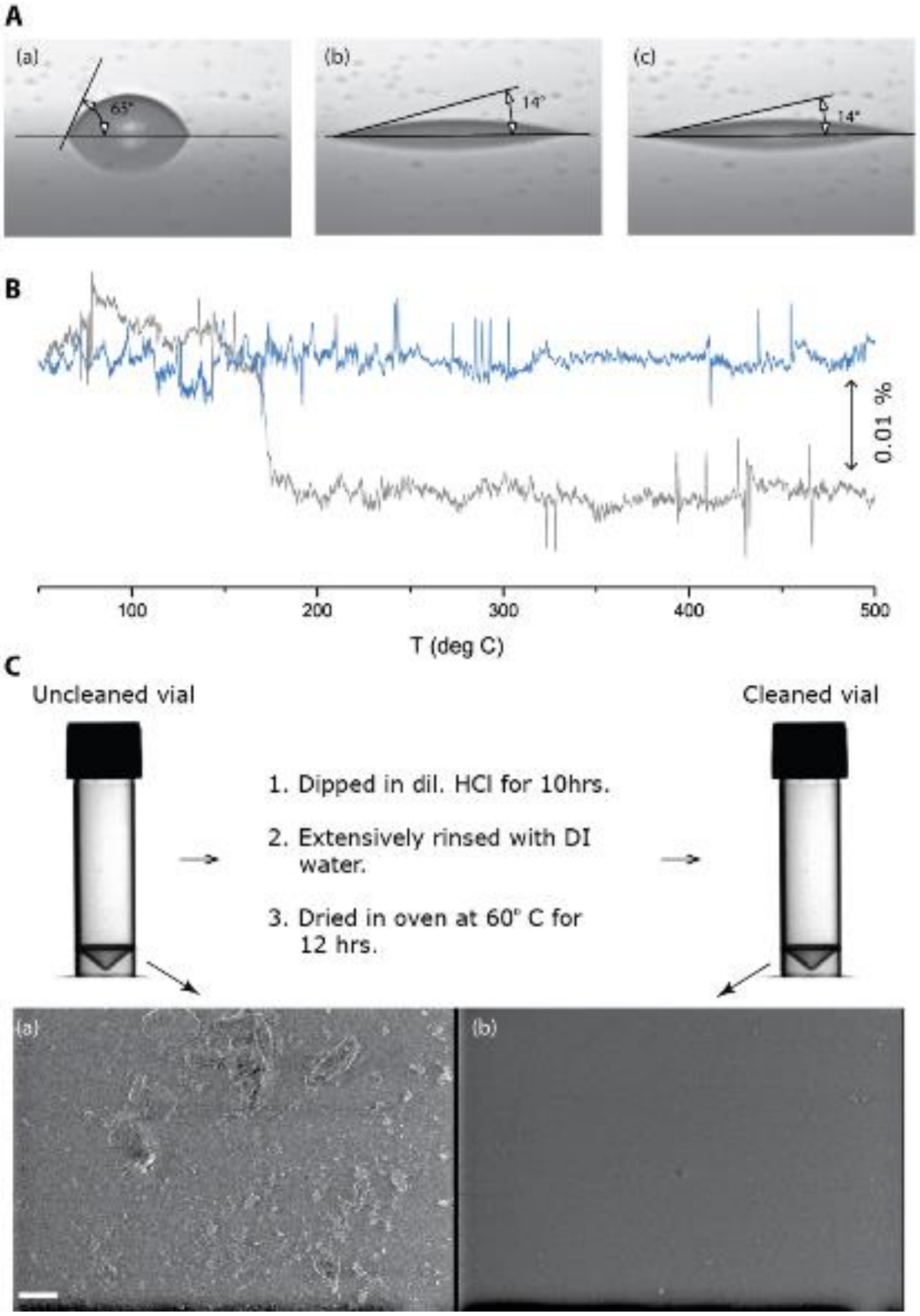

**Figure S2** Effect of cleaning glass particles and plastic vials. (A) The contact angle made by a drop of water on a glass slide (a) before and (b) after cleaning (see text for protocol) reduces from 65° to 14°. (c) Repeating the measurement on the same slide after 48 hours shows no significant change. (B) TGA measurement on uncleaned glass particles (gray line) shows an abrupt drop in mass by 0.01% at about $169° C$. Subsequent repeat run on the same sample (blue line) does not register the drop. (C) Scanning electron microscopy images of the surface of polypropylene vial (a) before and (b) after acid wash, showing that acid washing makes the surface smooth.

**Table T1 Effect of cleaning and environmental conditions**

| Humidity | Cleaning of glass particles | Cleaning of Polypropylene vial | Formation of regime ***I*** | Transition to regime ***II*** |
|---|---|---|---|---|
| >90 % | No | No | Yes | No |
| ~ 55 % | No | No | Yes | No |
| ~ 55 % | Heated at 200° C for 1 hr | No | Yes | No (see Fig.S5 A) |
| ~ 55 % | Heated at 200° C for 1 hr | Acid washed | Yes | Yes |
| > 90 % | Heated at 200° C for 1 hr | Acid washed | Yes | No |
| ~ 55 % | Heated @ 200° C for 1 hr | Acid washed for extended time | Yes | Yes. The transition is very abrupt. (see Fig.S5 B) |

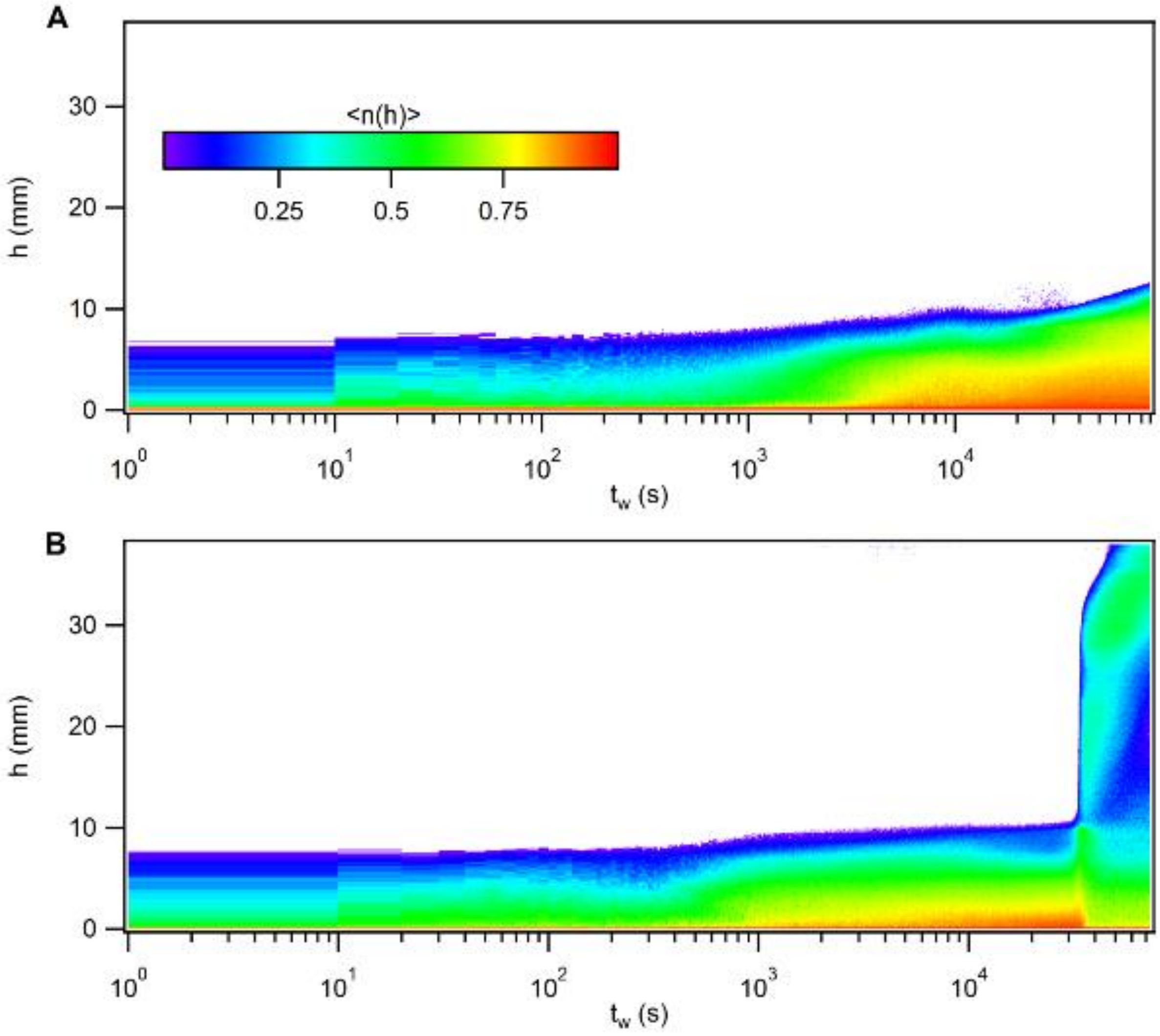

**Figure S3** The variation of $< \boldsymbol{n}(\boldsymbol{h}) >$ as a function of $\boldsymbol{t_w}$ shows that the transition to regime II depends surface conditions. (A) Uncleaned glass particles in uncleaned Polypropylene vial do not show transition into regime II even after ~20 hr. (B) For glass particles cleaned by heating as described in the text and for a vial that has been cleaned by a prolonged exposure (about 50 hr) to acid solution the transition to regime II happens in a more abrupt manner (compare with **Fig. 1C** of the paper).

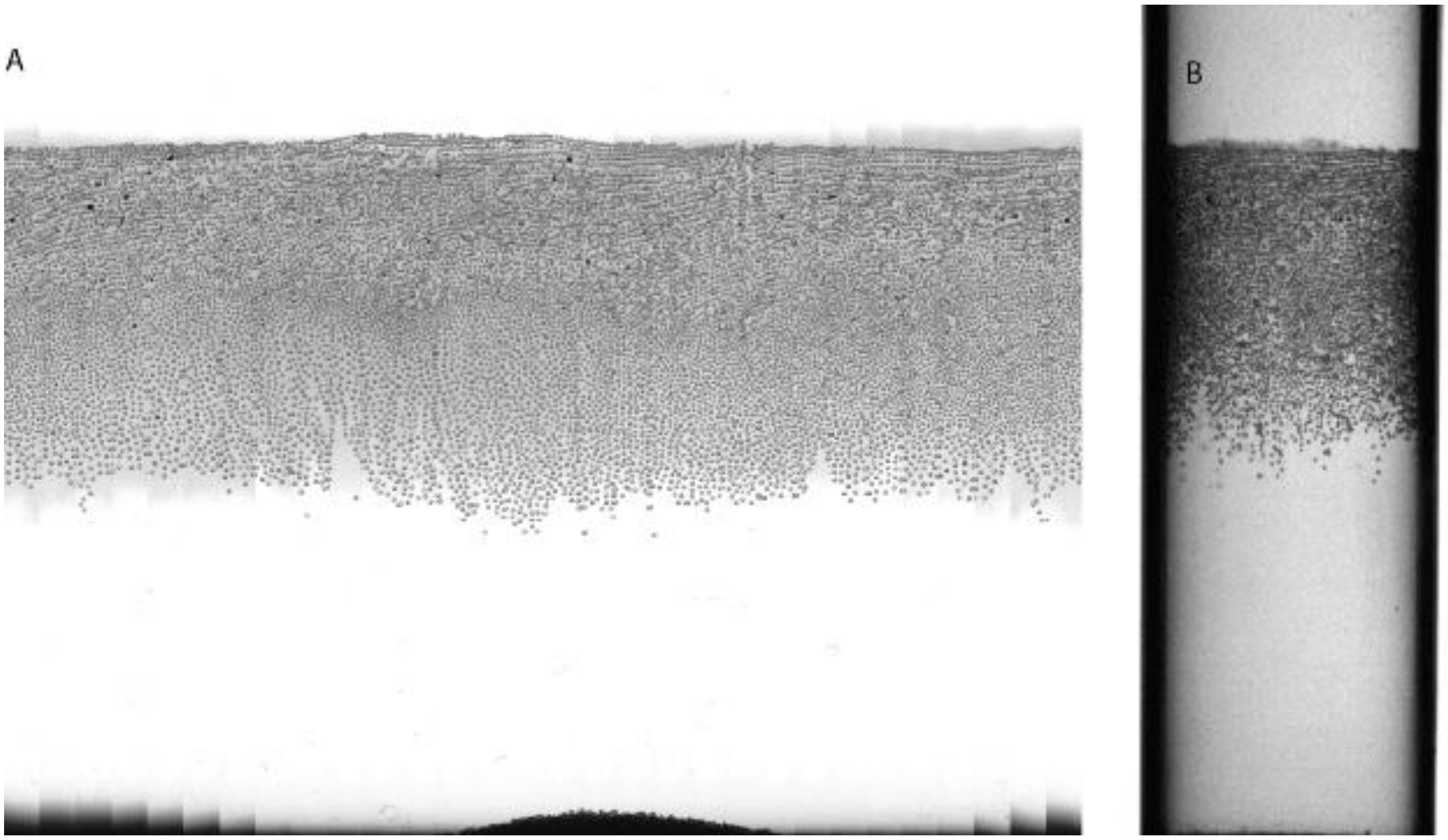

**Figure S4** (A) Stitched image of the entire cylindrical polypropylene surface obtained immediately after the completion of a typical experiment (52 hours of driving at $f = 14\ Hz$). (B) The same system imaged six months after the completion of the experiment.